# DISCRETE COSMOLOGICAL SELF-SIMILARITY

# AND DELTA SCUTI VARIABLE STARS

ROBERT L. OLDERSHAW

Amherst College

Amherst, MA 01002, USA

rloldershaw@amherst.edu

**Abstract.** Within the context of a fractal paradigm that emphasizes nature's well-stratified hierarchical organization, the δ Scuti class of variable stars is investigated for evidence of discrete cosmological self-similarity. Methods that were successfully applied to the RR Lyrae class of variable stars are used to identify Atomic Scale analogues of δ Scuti stars and their relevant range of energy levels. The mass, pulsation mode and fundamental oscillation period of a well-studied δ Scuti star are then shown to be quantitatively self-similar to the counterpart parameters of a uniquely identified Atomic Scale analogue. Several additional tests confirm the specificity of the discrete fractal relationship.

1. Introduction

In two previous papers (Oldershaw, 2008a,b) evidence was presented for discrete fractal phenomena associated with RR Lyrae variable stars. The masses, radii and oscillation periods of RR Lyrae stars were shown to have a discrete self-similar relationship to the masses, radii and transition periods of their Atomic Scale counterparts: helium atoms in moderately excited Rydberg states undergoing single-level transitions. The techniques that were used to achieve these unique results are applied here to a distinctly different class of variable stars: δ Scuti stars. The new results contain some interesting surprises, but they are in general agreement with the discrete fractal paradigm and offer additional evidence for the principle of discrete cosmological self-similarity.

Briefly, the Self-Similar Cosmological Paradigm (SSCP) proposes that nature is ordered in a transfinite hierarchy of discrete cosmological Scales, of which we can currently observe the Atomic, Stellar and Galactic Scales (Oldershaw, 1989a,b). Spatial lengths (R), temporal periods (T) and masses (M) of analogue systems on neighboring Scales $\Psi$ and $\Psi-1$ are related by the following set of discrete self-similar transformation equations:

$$R_\Psi = \Lambda R_{\Psi-1} \quad (1)$$

$$T_\Psi = \Lambda T_{\Psi-1} \quad (2)$$

$$M_\Psi = \Lambda^D M_{\Psi-1} \quad (3)$$

where Λ and D are dimensionless scaling constants with values of ≈ 5.2 x $10^{17}$ and ≈ 3.174, respectively, and $Λ^D$ ≈ 1.70 x $10^{56}$. The most readily available resource for a detailed presentation of this discrete fractal cosmology is the author's website (Oldershaw, 2001), where a full list of publications on the SSCP and downloadable copies of relevant papers are available. A familiarity with the previous fractal analysis (Oldershaw, 2008a) of the RR Lyrae class would be beneficial to a full appreciation of the present work on δ Scuti stars, but it is not mandatory.

## 2. Delta Scuti Variable Stars

Compared with RR Lyrae stars, δ Scutis are a somewhat erratic and heterogeneous class of stars. For example, the amplitudes of their oscillation periods can vary radically. As Breger and Pamyatnykh (2005) point out: "A star may change its pulsation spectrum to such an extent as to appear as a different star at different times", although "modes do not completely disappear, but are still present at small amplitudes." Here we will work exclusively with high amplitude δ Scuti stars (HADS) because their pulsation behavior is simpler (less multi-periodic) and more regular than low amplitude δ Scuti stars (LADS), and because they are thought to pulsate mainly in radial modes (Pigulski *et al.*, 2005) which is useful for identifying specific energy level transitions. Delta Scuti stars have spectral classifications of A to F and can be designated dwarf or subgiant stars (Alcock *et al.*, 2000). *Most importantly*, their masses typically range from 1.5 $M_⊙$ to 2.5 $M_⊙$ (Fox Machado *et al.*, 2005). This is a major change from the situation with the RR Lyrae class, which has a lower, narrower and better-defined mass range of 0.4 $M_⊙$ to 0.6 $M_⊙$.

As of 2004, roughly 400 δ Scuti stars were known. A typical oscillation period would be 0.1 day, and period cutoffs for this class occur at roughly 0.04 day and 0.2 day. Figure 1 shows a representative histogram of oscillation periods for 193 HADS (Pigulski et al., 2005). Superimposed upon this distribution are lines corresponding to typical oscillation periods for Stellar Scale n values, which will be derived and explained below.

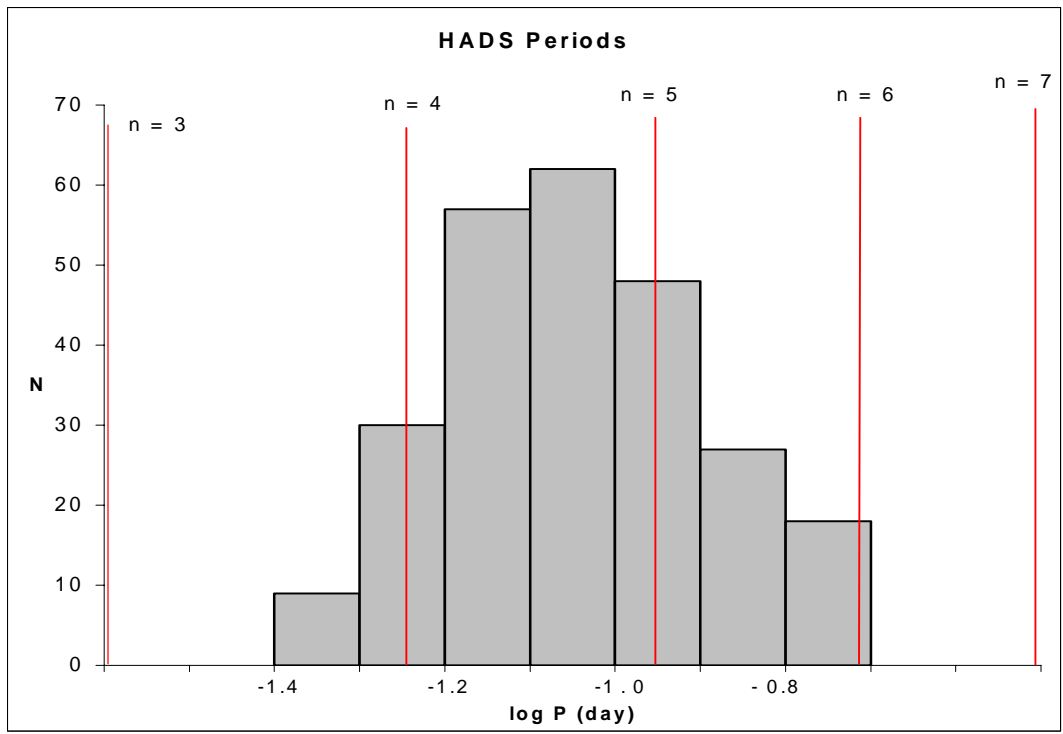

*Figure 1.* Histogram of periods for 193 high amplitude δ Scuti stars (Pigulski et al., 2005). Superimposed upon the period distribution are lines representing the *scaled* oscillation periods for Rydberg atoms excited to various n values.

In this particular sample of 193 HADS stars, there were 40 double-mode pulsators and twelve stars that simultaneously pulsed at 3 or more periods. Because data on the radii of HADS stars appear to be very limited, there is a minor problem with regard to identifying relevant energy levels, but the problem can be circumvented.

## 3. Atomic Scale Counterparts of Delta Scuti Stars

Given Eq. (3) and the mass of the proton ($m_p$), we can calculate that the Stellar Scale proton analogue will have a mass of about 0.145 $M_\odot$. The mass range (1.5 $M_\odot$ to 2.5 $M_\odot$) for δ Scuti stars would then correspond to an Atomic Scale mass range of approximately 10 $m_p$ to 17 $m_p$, or about 10 to 17 atomic mass units (amu). The atoms that dominate this mass range are Boron (11 amu), Carbon (12 amu), Nitrogen (14 amu) and Oxygen (16 amu). Therefore the majority of δ Scuti stars are hypothesized to be analogues of these atoms. Because reliable radius data are not available for δ Scuti stars, an alternative method that does not require $R_{\Psi=0}$ data must be used for determining the approximate energy levels of their relevant Atomic Scale analogue systems. There is a general relationship between the principal quantum numbers (n) for Rydberg atoms and their oscillation periods of the form:

$$p_n \approx n^3 p_0 \qquad (4)$$

where $p_0$ is the classical orbital period of the groundstate H atom: $\approx 1.5 \times 10^{-16}$ sec. The Stellar Scale equivalent to this relationship would be:

$$P_n \approx n^3 P_0 \qquad (5)$$

where $P_0 \approx \Lambda p_0 \approx (5.2 \times 10^{17})(1.5 \times 10^{-16} \text{ sec}) \approx 78$ sec. Using Eq. (5) we can generate the lines plotted in Fig. 1 for the different oscillation periods associated with Stellar Scale n values. These results tell us that if the discrete fractal paradigm is correct, and if we are dealing with analogues to Rydberg atoms undergoing transitions with $\Delta n \approx 1$, then the relevant range of n values for δ Scuti variables is predominantly $3 \leq n \leq 6$, with n = 5 being the most probable radial quantum number for this class of stars.

The above considerations lead us to conclude that δ Scuti variables correspond to a very heterogeneous set of B, C, N and O atoms excited to Rydberg states with n varying primarily between 3 and 6. The next step is to test this hypothesis by comparing the oscillation periods of δ Scuti stars with specific oscillation periods of their predicted Atomic Scale counterparts. However, there is a serious problem with a straightforward comparison between Stellar Scale and Atomic Scale frequency/period spectra, as was achieved in the case of the RR Lyrae class. The main source of this problem is the heterogeneity of the δ Scuti class of stars. We are dealing with analogues of *at least four different atoms*, each of which can be in several different isotopic configurations. Moreover, each of those four species of atom can be in several different ionization states: -1, neutral, +1, +2, etc. A further complication is that each of the atoms can have an entirely different set of energy levels for each of the singlet, doublet, triplet, etc., spin-related designations that apply. Also, we are less confident about using the $\Delta n \approx 1$ and $l \leq 1$ restrictions that were so helpful in the RR Lyrae case. Therefore, we are faced with an extremely large number of potential Atomic Scale transition periods to compare with the δ Scuti period distributions. A meaningful test requires that both the observed Stellar Scale period spectrum and the experimental Atomic Scale period spectrum have a limited number of uniquely identifiable peaks. In the case of the δ Scuti class of stars, the very large number of potential Atomic Scale comparison periods precludes a meaningful test of this type. Compounding this general lack of Atomic Scale specificity are the usual uncontrollable physical factors that can result in additional shifting of Stellar Scale energy levels away from unperturbed values: ambient Galactic Scale pressures, temperatures, electric fields and magnetic fields.

Fortunately, there is a way to circumvent the alarming specificity problems discussed above. If we have enough accurate information for an individual δ Scuti star, then we can use the data to identify the specific Atomic Scale analogue of the star, to restrict the number of possible energy level transitions for that specific atom, and to construct a valid test between a uniquely predicted oscillation period and a limited number of Atomic Scale comparison periods. Very recently, the δ Scuti star GSC 00144–03031 was analyzed in detail by Poretti *et al* (2005) and this system has certain characteristics that make it an excellent test star for our purposes. First and foremost, its mass has been determined with reasonable accuracy and is approximately 1.75 $M_\odot$. Using our knowledge that 0.145 $M_\odot \approx$ 1 smu (stellar mass unit, $\approx \Lambda^D$ amu), we can determine that 1.75 $M_\odot$ corresponds to 12 smu and therefore GSC 00144–03031 can be identified with a high degree of confidence as an analogue of a $^{12}C$ atom. Other advantages of using this star as a test system are that it is a classic HADS system (regular, high amplitude pulsations), that it has a dominant fundamental mode that is highly radial in character (which helps in narrowing energy level possibilities), and finally that it is a pure double-mode pulsator (providing us with an second test period). The fundamental radial mode has a period of about 0.058 day ($\approx$ 5017.42 sec) and its amplitude is 4 times greater than the secondary pulsation which has a period of ≈ 3872.70 sec. The basic characteristics of GSC 00144-03031 are summarized in Table 1.

*Table 1.* Physical Properties (Poretti *et al.*, 2005) of GSC 00144–03031

| **Class** | high amplitude δ Scuti (HADS) |
|---|---|
| **Mass** | ≈ 1.75 $M_\odot$ |
| **Mode** | "pure double-mode pulsator" with radial fundamental mode |
| **Fundamental Period** | 5017.42 sec (radial, amplitude = 0.1383 mag) |
| **Secondary Period** | 3872.70 sec (amplitude = 0.0331mag) |

### 4. Test of the Discrete Self-Similarity Principle

If the principle of discrete cosmological self-similarity is correct, then we should find a self-similar relationship between the oscillation periods of GSC 00144-03031 and the empirical oscillation periods of its Atomic Scale analogue undergoing corresponding transitions, in accordance with Eq. 2. We have identified the Atomic Scale analogue as a $^{12}C$ atom and the transition has a strong radial mode (l = 0) character. From Figure 1, we can determine that the position of the dominant period for GSC 0144–03031 falls between the n ≈ 5 and the n ≈ 4 lines and so we anticipate a correlation with a n = 5 → 4, low l, transition. Using Eq. 2 we can calculate a predicted Atomic Scale period for the counterpart to the δ Scuti fundamental mode:

$$P_{\Psi-1} \approx P_\Psi \div \Lambda \approx 5017.42 \text{ sec} \div 5.2 \times 10^{17}$$

$$\approx 9.65 \times 10^{-15} \text{ sec.}$$

We assume that the $^{12}$C atom is most likely to be uncharged, rather than being in an ionized state. Therefore the quantitative test is whether a neutral $^{12}$C atom has a radial mode transition between the n = 5 and n = 4 levels that involves an oscillation period of about 9.65 x $10^{-15}$ sec. We use a standard source for atomic energy level data (Bashkin and Stoner, Jr, 1975) and find that the n = 5 energy level with the least non-radial character is the $1s^22s^22p5p$(J=0) $^1$S singlet level with an energy of 85625.18 cm$^{-1}$. The n = 4 energy level with the least non-radial character is the $1s^22s^22p4p$(J=0) $^1$S singlet level with an energy of 82251.71 cm$^{-1}$. Subtracting the energies for these neighboring energy levels gives 3373.47 cm$^{-1}$ as the transition energy ($\Delta$E) for the n = 5 → 4(J=0) transition. We can calculate the oscillation frequency for the transition by using the relation $\nu = \Delta E c$ for electromagnetic radiation, and we find that $\nu$ = (3373.47 cm$^{-1}$)(2.99 x $10^{10}$ cm/sec) = 1.01 x $10^{14}$ sec$^{-1}$. Since p = 1/$\nu$, the period for the transition is 9.88 x $10^{-15}$ sec. This value is higher than the predicted value of 9.65 x $10^{-15}$ sec by a factor of 0.024, but considering the numerous sources of small uncertainties that are involved in this test, and the uncontrollable physical factors that can shift the Stellar Scale oscillation period, the agreement between the predicted and experimental values is quite good. Table 2 summarizes the discrete self-similarity between GSC 00144–03031 and $^{12}$C [$1s^22s^22p5p$ → 4p, (J=0), $^1$S].

*Table 2.* Comparison of the Fundamental Mode Properties of GSC 00144–03031 and Its $^{12}$C Analogue Undergoing a [$1s^22s^22p5p(J=0) \to 4p(J=0)$, $^1$S] Transition

| Parameter | GSC 00144-03031 | Scale Factor | Predicted Analogue Values | Empirical $^{12}$C values | Error |
|---|---|---|---|---|---|
| **Mass** | $\approx 1.75\ M_\odot$ | $1/\Lambda^D$ | $\approx 12$ amu | $\approx 12$ amu | - |
| **Fund. Mode** | radial | - | radial | $\approx$ radial | 0 |
| **n** | $4 \leq n \leq 5$ | - | $4 \leq n \leq 5$ | $4 \leq n \leq 5$ | 0 |
| **Period** | 5017.42 sec | $1/\Lambda$ | $9.65 \times 10^{-15}$ sec | $9.88 \times 10^{-15}$ sec | 0.024 |

To verify the uniqueness of the discrete self-similar relationship between the specific oscillation periods of GSC 00144–03031 and $^{12}$C [$5p(J=0) \to 4p(J=0)$, $^1$S], the oscillation periods of other transitions in the $3 \leq n \leq 5$ range were checked. The closest alternative match occurred for the [$5p(J=2) \to 4p(J=2)$, $^1$D] transition. However, its oscillation period of $9.18 \times 10^{-15}$ sec is about 5% low and the transition is <u>not</u> similar to a fundamental radial mode oscillation. Oscillation periods for other $^{12}$C transitions [$3 \leq n \leq 5$; $^1$S and $^3$S] differed from the predicted period of $9.65 \times 10^{-15}$ sec by 10% or more. Given the good quantitative match between the fundamental period of GSC 00144–03031 and the *single uniquely specified* transition period of $^{12}$C, it seems likely that discrete cosmological self-similarity has been shown to apply in this case.

To further demonstrate the uniqueness of our result, we can repeat the same analysis for other atoms such as H, He, Li, Be, B and N. The results of these calculations are summarized in Table 3.

*Table 3.* Oscillation Periods [5(l ≈ 0) → 4(l ≈ 0)] for Atoms Other Than $^{12}$C

| Atom | ΔE (cm$^{-1}$) | P (sec) |
|---|---|---|
| H | 2467.78 | 1.35 x 10$^{-14}$ |
| He | 2723.28 | 1.22 x 10$^{-14}$ |
| Li | 3287.47 | 1.02 x 10$^{-14}$ |
| Be | 4076.88 | 8.18 x 10$^{-15}$ |
| B | 5136.47 | 6.49 x 10$^{-15}$ |
| N | 4583.11 | 7.27 x 10$^{-15}$ |

The oscillation periods for the most radial transitions [5(l=0) → 4(l=0)] of H, He, Be, B and N are definitely *not* in good agreement with our predicted test period. The closest alternative match is the 5s → 4s transition for Li with an oscillation period of 1.02 x 10$^{-14}$ sec, which is higher than our predicted period by about 5.4%, and would require an unreasonable 42% error in the mass estimate for GSC 00144–03031.

## 5. The Secondary Period of GSC 00144–03031

As an additional check on the uniqueness of the above results, we now explore the secondary oscillation period of GSC 00144–03031, which is 3872.70 sec. At this point in the development of the SSCP, we are still trying to fully understand single-mode pulsators, although research on double-mode pulsation is on-going and seems promising. That preliminary work suggests that *both* oscillation periods of a double-mode pulsator come from the discrete spectrum of allowed transition periods for that system. Therefore, we can predict that for a $^{12}$C atom undergoing transitions with 3 ≤ n ≤ 5, there will be a transition period very close to (3873 sec)(1/Λ) ≈ 7.45 x 10$^{-15}$ sec. Actually, when this prediction is tested we find three candidates. The [5p(J=1) → 4p(J=1), $^1$P] transition comes within 4.4% of the predicted period, but the [(J=1), $^1$P] nature of this transition does not have much radial character and we expect the match between predicted period and comparison period to be at the 3% level, or better. The triplet configuration of $^{12}$C has a [$1s^22s^22p(^2P^o)5s$(J=2) $\rightarrow$ $1s^22s^22p3d$(J=1), $^3P^o$] transition with a period of 7.48 x 10$^{-15}$ sec (only 0.4% high), and the (J=0) version of that transition has a period of 7.55 x 10$^{-15}$ sec (1.3% high). These two closely related transitions offer very good quantitative matches, but the acceptability of having simultaneous transitions involving *both* singlet and triplet configurations remains to be more fully explored. The third potential match, and possibly the most interesting, occurs with the [$1s^22s^22p4p$(J=0), $^1$S $\rightarrow$ $1s^22s^22p3d$(J=2), $^1D^o$] transition. This transition has an oscillation period of 7.32 x 10$^{-15}$ sec, which is quite close to the predicted period of 7.45 x 10$^{-15}$ sec (1.8% low). It also has the unique feature that it is directly linked to the fundamental pulsation period since both transitions share the [$1s^22s^22p4p$(J=0), $^1$S] energy level. Whereas the previous candidates would seem to require some sort of "superposition" of possibly competing transitions, the third candidate suggests an alternative qualitative explanation for double-mode pulsation, wherein the system is undergoing a sequence of two separate, but related, transitions and the second oscillation begins to activate before the first oscillation has completely finished.

    Although at present we do not have enough information to definitively choose between the three candidate matches for the secondary oscillation of GSC 00144–03031, we can safely say that this additional check on the uniqueness of the primary results for the dominant oscillation period has yielded encouraging results. Had we not found any

period matches at the < 5% level, then that might have indicated a serious problem with the analysis, or possibly with the whole concept of discrete cosmological self-similarity.

### 6. Conclusions

The δ Scuti class of variable stars is a much more heterogeneous class than the RR Lyrae class, corresponding to a collection of Atomic Scale systems with masses in the 10 to 17 amu range. The *high amplitude* δ Scuti stars appear to be limited to low $\Delta n$ transitions ($\Delta n \approx 1$) primarily within the range $3 \leq n \leq 6$. The substantial heterogeneity of this class interferes with a simple comparison of sizeable samples of empirical δ Scuti oscillation periods with predicted periods derived from Atomic Scale data, although this may be possible *in principle*. However, we have achieved the specificity required for a meaningful test of discrete cosmological self-similarity by focusing on an individual, well-characterized δ Scuti star. Based purely on physical data for GSC 00144–03031, we have identified:

(1) a specific Atomic Scale analogue ($^{12}C$),

(2) a most likely energy level transition ($1s^2 2s^2 2p5p \to 4p$, J=0, $^1S$), and

(3) a uniquely matching self-similar oscillation period (agreement at the 97.6% level).

These new results, combined with the previous successful demonstration (Oldershaw, 2008a,b) of discrete self-similarity between RR Lyrae stars and He atoms undergoing $\Delta n = 1$ Rydberg state transitions, lend further support to our contention that discrete cosmological self-similarity is a fundamental property of nature. It can be predicted that the same methods that have been applied here, and in the case of the RR Lyrae stars, can be successfully applied to other δ Scuti stars if the following criteria are met. The stellar mass <u>must</u> be known to an accuracy of ≤ 0.05 $M_\odot$, so that the correct Atomic Scale analogue can be identified. Ideally the star should pulsate in a single dominant oscillation mode, although double-mode pulsators can also be analyzed by our methods. Multi-mode pulsators with three or more low-amplitude periods appear to be analogous to excited, highly perturbed, atomic systems that are oscillating at several *potential* transition periods, but are not yet undergoing single specific transitions between energy levels, as will be discussed in a forthcoming paper (Oldershaw, 2008c) on the class of ZZ Ceti variable stars. At any rate, the higher the amplitude of the dominant oscillation period of the δ Scuti star, the more likely it is that we are observing an event that is self-similar to a full-fledged transition between discrete energy levels. Although we may be getting a bit ahead of ourselves here, it is conceivable that a typical single-mode HADS star evolves from a multi-mode LADS star when the latter absorbs a sufficient amount of energy at an appropriate frequency in order to trigger a genuine energy level transition.